\documentclass[a4paper]{jpconf}
\usepackage{graphicx}
\usepackage{hyperref}
\usepackage{amsmath,amssymb}
\usepackage{color}
\usepackage[caption=false]{subfig}
\def\eq#1{(\ref{#1})}

\graphicspath{{./figures/}}

\hypersetup{
	pdftitle={Phase Structure and Dynamics of QCD-- A Functional Perspective},
	pdfauthor={N. Strodthoff},
	pdfkeywords={QCD} {Yang-Mills theory} {Spectral Functions} {transport coefficients} {FRG},
	bookmarksopen=false,
	bookmarksnumbered=true
}

\begin{document}
\title{Phase Structure and Dynamics of QCD--\\A Functional Perspective}

\author{Nils Strodthoff}

\address{Nuclear Science Division, Lawrence Berkeley National Laboratory, Berkeley, CA 94720, USA}

\ead{nstrodthoff@lbl.gov}

\begin{abstract}
The understanding of the phase structure and the fundamental properties of QCD matter from its microscopic description
requires appropriate first-principle approaches. Here I review the progress towards a quantitative
first-principle continuum approach within the framework of the Functional Renormalization group established by the fQCD collaboration. I
focus on recent quantitative results for quenched QCD and Yang-Mills theory in the vacuum before addressing the calculation of dynamical quantities
such as spectral functions and transport coefficients in this framework.
\end{abstract}

\section{Introduction}
The understanding of the phase structure of QCD from first principles has been on the forefront of research in hadron physics for several decades \cite{Fukushima:2010bq}. While Lattice QCD provides a reliable approach at vanishing and small chemical potentials, the fermion sign problem makes reliable predictions
at large chemical potentials, $\mu_B\gtrsim T$, very difficult \cite{deForcrand:2010ys}. The latter are however required to make quantitative predictions about
the existence and location of a possible critical point in the phase diagram, a question which has been one of the driving forces for theoretical as well as experimental efforts in the field. The second major challenge is to gain a quantitative understanding of the fundamental properties of QCD matter from a microscopic description. This includes observables connected to the hadron spectrum such as pole masses,
decay constants, form factors or scattering amplitudes but also real-time observables like elementary spectral functions and transport coefficients with particular relevance for phenomenological applications. Such real-time observables are notoriously difficult to obtain in Euclidean approaches due to the necessity of performing an analytic continuation from Euclidean to Minkowski signature. 

Both fundamental challenges require first-principle approaches to QCD in order to make quantitative predictions. The fQCD collaboration \cite{fQCD:2016-10} aims to provide such a quantitative continuum approach to QCD, cf.\ \cite{Pawlowski:2014aha} for an overview, \cite{Mitter:2014wpa,Cyrol:2016tym} for recent quantitative works in quenched QCD and Yang-Mills (YM) theory and \cite{Braun:2014ata} for an exploratory study of the unquenched system. 
This implies in particular to renounce the use of phenomenological input and to restrict the input parameters to those of QCD itself, namely the strong coupling and the quark mass at perturbative momentum scales.

\section{Euclidean correlation functions from continuum QCD}
Functional continuum approaches exploit relations between off-shell Green's functions. In the following I will focus on the Functional Renormalization group (FRG) which represents a powerful non-perturbative tool that implements the idea of the Wilson RG of gradually integrating quantum fluctuations momentum shell by momentum shell in a formulation in momentum space, see \cite{Berges:2000ew,Pawlowski:2005xe,Gies:2006wv,Schaefer:2006sr,Braun:2011pp} for QCD-related reviews. Technically this is achieved by adding an infrared regulator to the theory which acts like a mass term thereby suppressing fluctuations below a certain cutoff scale. Consequently all objects in the theory carry a dependence on this artificial regulator scale. The central object in this approach is the scale-dependent analogue of the effective action $\Gamma_k$ whose scale dependence is governed by a one loop equation,
\begin{equation}
\label{eq:flow}
k\partial_k \Gamma_k=\frac{1}{2}\text{Tr}\frac{1}{\Gamma^{(2)}_k+R_k}\,k\partial_k R_k\,,
\end{equation}
where $\Gamma^{(2)}_k$ denotes the full scale-, momentum- and field-dependent inverse propagator and $R_k$ is the regulator term. Flow equations for higher $n$-point function can then be derived by taking functional
derivatives with respect to the fields in \eq{eq:flow}. These coupled systems of functional equations then have to be integrated from some large UV scale down to $k=0$ where the regulator term is removed
and the effective action is recovered. The practical application requires the use of appropriate non-perturbative expansion schemes to
render the infinite tower of coupled equations tractable. One example for such a scheme is the vertex expansion corresponding to an expansion of the effective action in terms of 1-particle irreducible vertex functions.

\begin{figure*}[t]
  \centering \subfloat[Gluon propagator in Yang-Mills theory in comparison to lattice results (black dots); figure taken from \cite{Cyrol:2016tym}. \hfill\textcolor{white}{.}]{\includegraphics[width=0.47\textwidth]{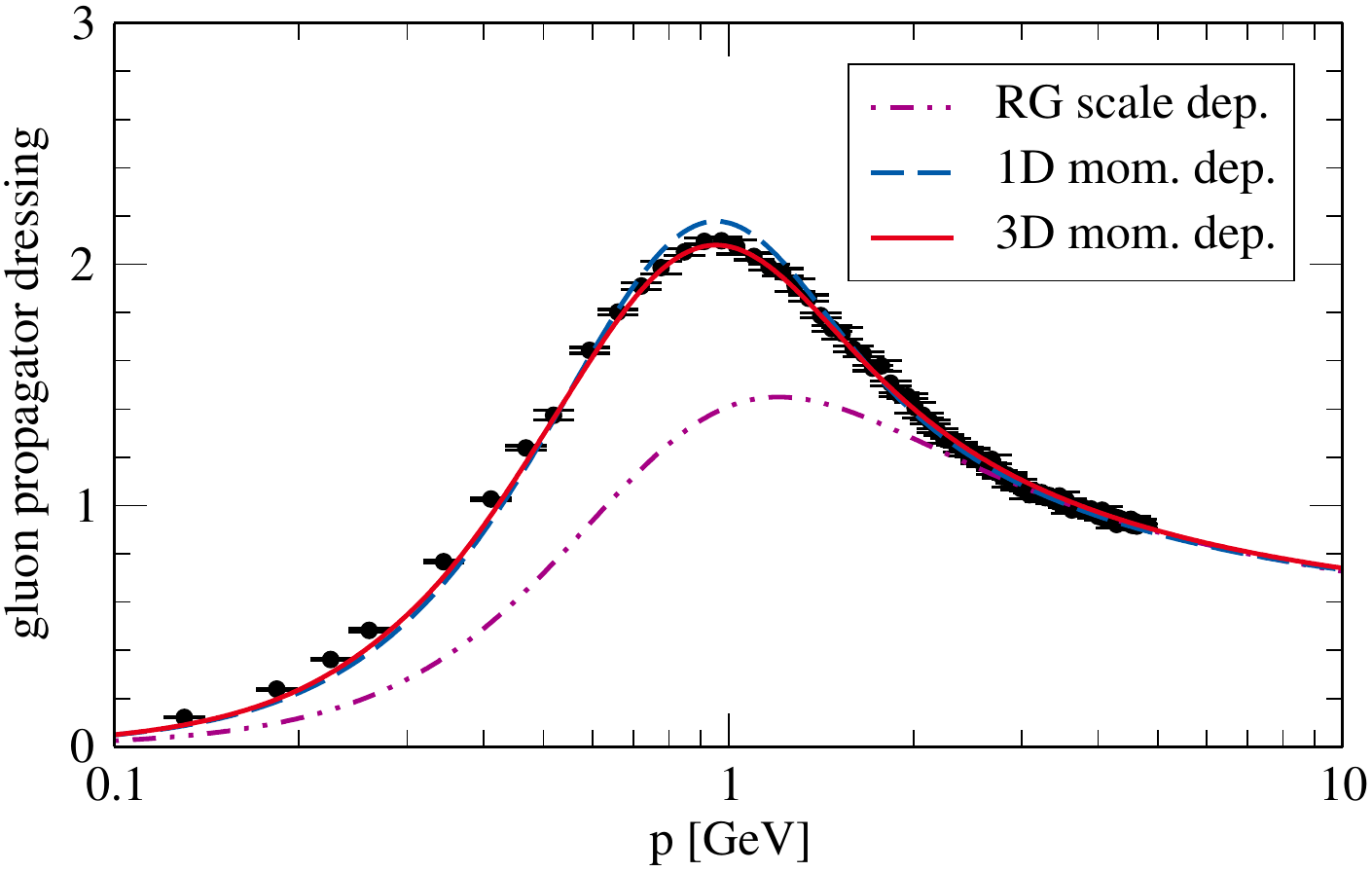}\label{fig:gluonprop}}\hfill \subfloat[Quark propagator in quenched QCD in comparison to lattice results; figure taken from \cite{Mitter:2014wpa}.\hfill\textcolor{white}{.}]{
\includegraphics[width=0.47\textwidth]{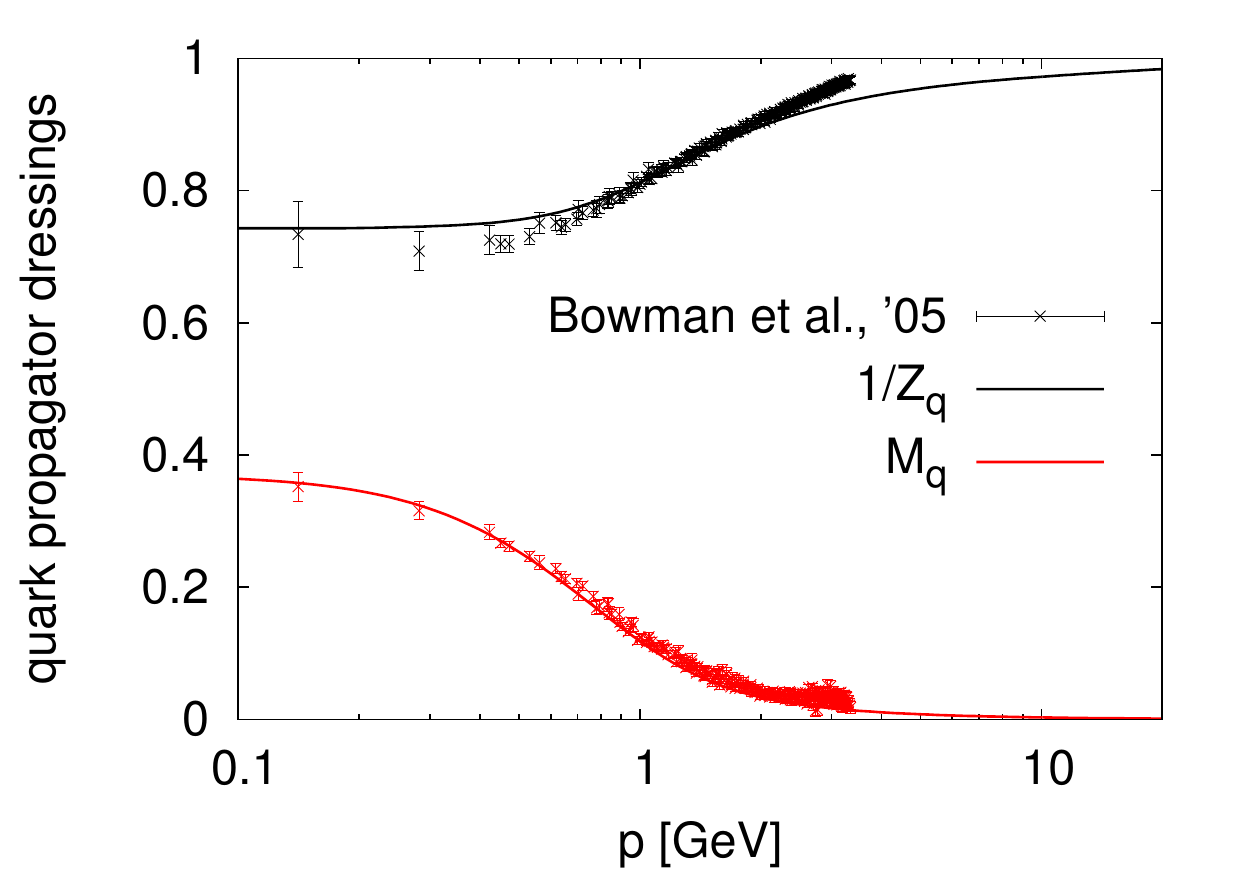}
	\label{fig:quarkprop}}
\caption{Euclidean propagator results from first-principle continuum QCD\hfill\textcolor{white}{.} }\label{fig:Euclprop}
\end{figure*}

This was put into practice in a calculation of the transverse correlation functions in Landau-gauge Yang-Mills theory \cite{Cyrol:2016tym} by self-consistently solving the system
of propagators and primitively divergent vertex functions. The resulting gluon propagator is found to be in good agreement with lattice results, see Fig.~\ref{fig:gluonprop}. 
This study is complemented by a study of the quenched matter system using only gluon and ghost propagators as input \cite{Mitter:2014wpa}. Again
the quark propagator is found in very good agreement with the lattice, see Fig.~\ref{fig:quarkprop}. To achieve this
quantitative agreement it is required to consider a large truncation scheme including in particular a full tensor basis for the quark-gluon vertex as well as non-classical higher-order
quark-gluon interaction vertices, which is only possible with the help of appropriate (computer-)algebraic tools \cite{Huber:2011qr,Cyrol:2016zqb}. The emergence of effective mesonic degrees of freedom is captured by means of the dynamical hadronisation technique, see \cite{Gies:2001nw} and \cite{Mitter:2014wpa} for the momentum-dependent generalization, which recasts resonant structures in four-fermi interactions in terms of effective mesonic interactions. 

First quantitative results for the coupled system of QCD correlation functions in the vacuum are currently in preparation and will soon be extended to finite temperature. The further extension to moderately large
chemical potentials will then allow for example studies of fluctuation observables, see \cite{Fu:2016tey} for a recent model study in this direction.

\section{Direct calculation of real-time observables}
\begin{figure*}[t]
  \centering \subfloat[Self-consistent spectral functions in the $O(N)$ model; figure adapted from \cite{Strodthoff:2016pxx}.\hfill\textcolor{white}{.}]{
\includegraphics[width=0.47\textwidth]{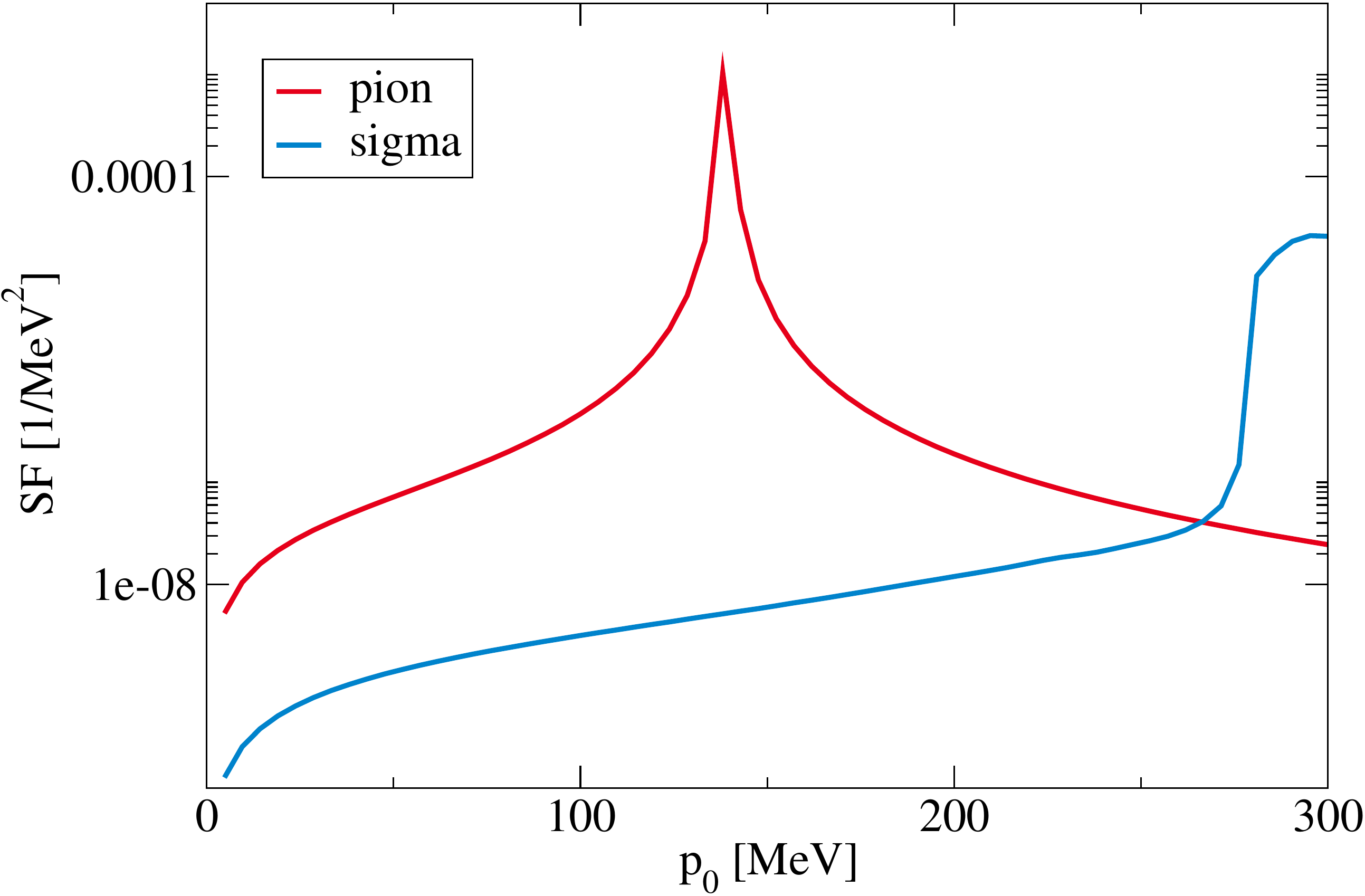}
	\label{fig:spectralon}} \hfill \subfloat[Shear viscosity over entropy density ratio $\eta/s$ in YM theory; figure taken from \cite{Christiansen:2014ypa}. \hfill\textcolor{white}{.}]{\includegraphics[width=0.47\textwidth]{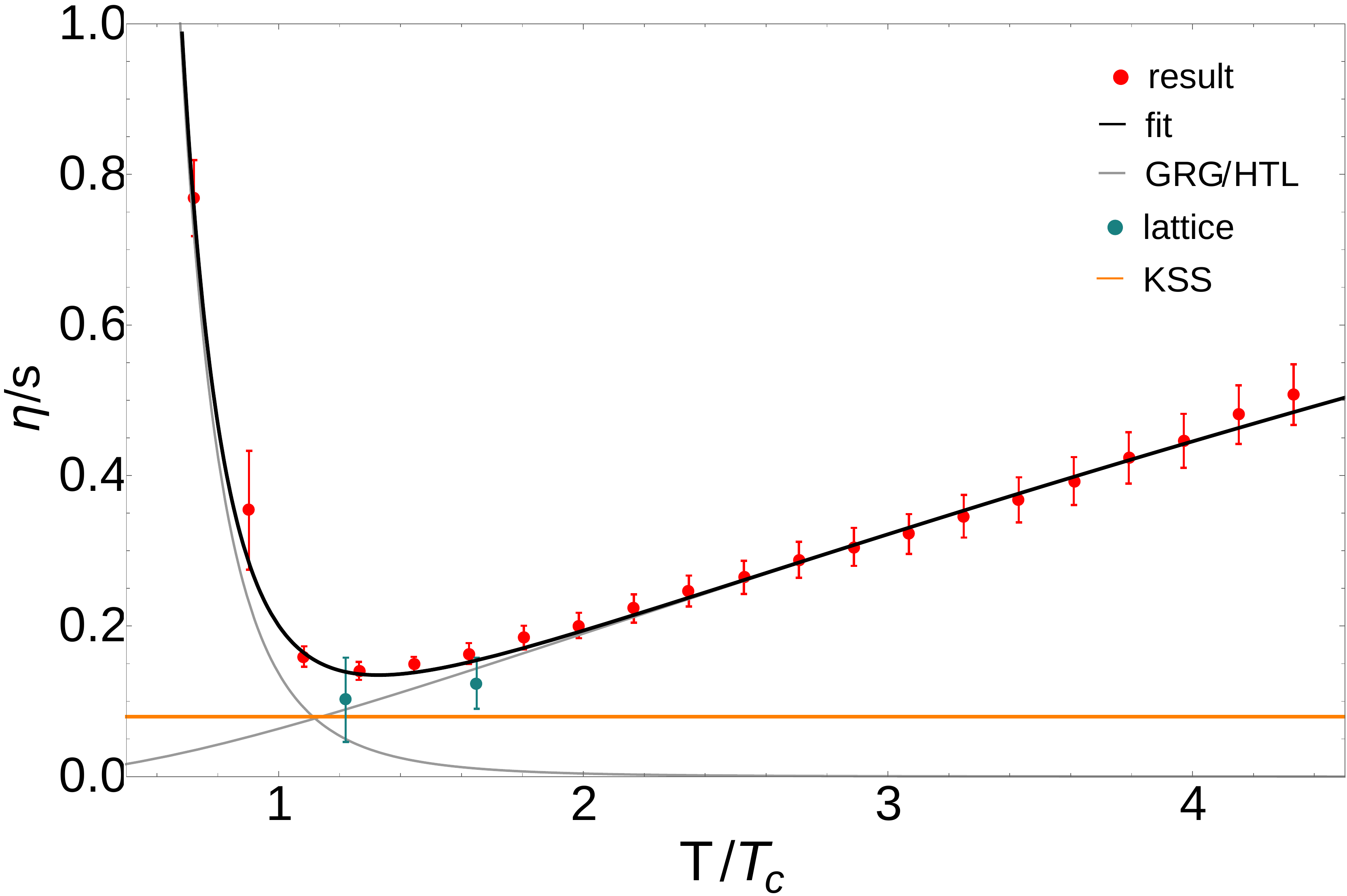}
\label{fig:etas}}
\caption{Exemplary real-time results from functional approaches\hfill\textcolor{white}{.} }\label{fig:real-timeresults}
\end{figure*}
Apart from the promising self-consistent solutions of Euclidean correlation functions, there has been tremendous progress in the direct calculation of real-time observables within the FRG. The central difficulty in calculating real-time observables from a Euclidean approach is the necessity of performing an analytic continuation in the external momentum. The standard procedure is to use Euclidean data and to infer the analytic continuation numerically e.g.\ by means of the Maximum Entropy Method (MEM). However, this procedure goes along with large systematic errors, which urges for a direct calculation of real-time observables. 

The main idea of the approach initiated in \cite{Kamikado:2013sia} is the fact that the external momentum appears, very much like the chemical potential, as an external parameter in the functional equations. This allows an analytic continuation on the level of the equation itself and hence a direct calculation of elementary spectral functions anchored to a quantitative Euclidean baseline. This procedure has been applied successfully in scalar models \cite{Kamikado:2013sia} and Yukawa systems \cite{Tripolt:2013jra,Yokota:2016tip,Jung:2016yxl}. The application to numerical propagators as prerequisite for studies of spectral functions in full QCD required a significant extension of the original formalism \cite{Pawlowski:2015mia}. The applicability of this procedure to include also momentum-dependent vertices and the full (complex) momentum dependence has recently been demonstrated in \cite{Strodthoff:2016pxx}
at the example of the first self-consistent direct calculation of spectral functions in the $O(N)$ model, see Fig.~\ref{fig:spectralon}.

Elementary spectral functions do not only provide physical insights by itself but furthermore provide input for the calculation of transport coefficients by means of Kubo formulae, which relate the latter to
correlation functions of the energy-momentum tensor. These can in turn be calculated within a finite diagrammatic expansion in terms of full propagators and vertices, which represent the only input of the calculation.
This procedure has been applied to the calculation of the shear viscosity over entropy density ratio $\eta/s$ in Yang-Mills theory \cite{Haas:2013hpa} using gluonic spectral functions obtained via MEM as input. This allows to determine the temperature dependence of $\eta/s$ over wide range of temperatures, see Fig.~\ref{fig:etas}. The asymptotic behavior is found to be in agreement with perturbative results whereas
the low temperature regime is well-described by a power-law decay attributed to a glueball resonance gas similar to the hadron resonance gas in QCD. Based on a direct sum of these two contributions we provided a first estimate for $\eta/s$ in QCD \cite{Haas:2013hpa}. This represents the first step towards a direct determination of transport coefficients from the microscopic dynamics as urgently required input for hydrodynamical calculations.

\section{Summary and Outlook}
In this contribution I reviewed the recent progress on the QCD phase diagram and the direct calculation of real-time observables within the framework of the FRG. Firstly, there are by now quantitative results
available for Yang-Mills theory and QCD with extensions to finite temperature and density on the way. These were obtained by self-consistently solving the systems of correlation functions renouncing further
phenomenological input. The extension to finite chemical potential will allow for example the calculation of fluctuation observables such as cumulants of the net-baryon distribution or couplings \cite{Bzdak:2016sxg} possibly coupled to the hydrodynamic evolution and including non-equilibrium effects \cite{Herold:2016uvv}. Secondly, there has been tremendous progress in a direct calculation of real-time observables in this framework. In combination with the quantitative Euclidean results from above this approach has very bright prospects for a direct calculation of elementary spectral functions which can then in a second step be used for the calculation of transport coefficients in full QCD. 
\vspace{-2ex}
\section*{Acknowledgments}
The author thanks the fQCD collaboration and in particular N.~Christiansen, A.~K.~Cyrol, M.~Mitter and J.~M.~Pawlowski for joint work on topics discussed above. This work was
supported by the DFG under grant no.\ Str1462/1-1 and by the Office of Nuclear Physics in the US Department of Energy's Office of Science under Contract No. DE-AC02-05CH11231.
\vspace{-2ex}
\section*{References}
\bibliographystyle{iopart-num}
\bibliography{../bib_master}
\end{document}